\documentclass[%
 reprint,
 amsmath,amssymb,
 aps,prl
]{revtex4-2}

\usepackage{graphicx}% Include figure files
\usepackage{dcolumn}% Align table columns on decimal point
\usepackage{bm}% bold math
\usepackage[separate-uncertainty = true]{siunitx}

\begin{document}

\title{Channel closing in strong-field photoemission from tip arrays}% Force line breaks with \\

\author{Leon Brückner}
\email{leon.brueckner@fau.de}
\affiliation{Department of Physics, Friedrich-Alexander-Universität Erlangen-Nürnberg, 91058 Erlangen, Germany}
\author{Jonas Heimerl}%
\affiliation{Department of Physics, Friedrich-Alexander-Universität Erlangen-Nürnberg, 91058 Erlangen, Germany}
\author{Constantin Nauk}
\altaffiliation{Present address: Physikalisch-Technische
Bundesanstalt, 38116 Braunschweig, Germany}
\affiliation{Department of Physics, Friedrich-Alexander-Universität Erlangen-Nürnberg, 91058 Erlangen, Germany}
\author{Peter Hommelhoff}
\email{peter.hommelhoff@lmu.de}
\affiliation{Department of Physics, Friedrich-Alexander-Universität Erlangen-Nürnberg, 91058 Erlangen, Germany}
\affiliation{Fakultät für Physik, Ludwig-Maximilians-Universität, 80539 München, Germany}

\date{\today}

\begin{abstract}
Ultrafast photoemission is a promising avenue for the development of petahertz electronics. For such devices, identifying the precise emission mechanism and quantifying the local field strength at the emission site are of utmost interest. While this information can be extracted from electron energy spectra or the scaling of the emission rate with peak intensity, such measurements are not always technically feasible and a clear interpretation of the results can be challenging. Here, we demonstrate channel closing as a precision gauge for the local enhanced optical nearfield strength. Channel closings originate from a field-induced upshift of the vacuum level, which leads to a small decrease of the emission yield with increasing intensity. We report the first experimental observation of such channel closings in the photoemission rate from an array of nanometer-sharp gold tips. Our data are well matched by simulated yields, enabling us to precisely measure the local, field-enhanced optical nearfield at the tip array \textit{in situ}.
We expect this method to be a versatile diagnostic tool for the characterization of lightwave optoelectronic devices, as it suffices to measure currents only.
\end{abstract}

\maketitle
\noindent Understanding the electron emission dynamics from atoms, molecules, and solids under the influence of extreme laser fields is at the core of strong-field physics \cite{Corkum1993,Krausz2009,Ciappina2017,Lhuillier2024}. More recently, these ultrafast electron dynamics have been applied to ultrafast signal processing in nano-fabricated devices, called lightwave or petahertz electronics \cite{Borsch2023,Heide2024}. In petahertz electronic devices based on ultrafast photoemission, the ultrafast photoemission typically serves as a (sub-) femtosecond-fast switch, facilitating the potentially petahertz-scale bandwidths sought after. Achieving these bandwidths requires operating within the tunneling regime of photoemission \cite{Krüger2011,Herink2012,Rybka2016,Karnetzky2018,Bionta2021}. In this regime, the photoemission rate varies largely as a function of the electric field strength and contains deep insights to the underlying physics.
Common theory models that describe this emission process include the Keldysh rate \cite{Keldysh1964}, the ADK-rate \cite{Ammosov1986}, and the Yudin-Ivanov rate \cite{Yudin2001}. For a single active electron, the electron yield is predicted to scale as a power law at low intensities ($\gamma \gg 1$), where $\gamma=\sqrt{\Phi / 2U_p}$ denotes the Keldysh parameter, with the work function $\Phi$ and the ponderomotive energy $U_p$. In this case, the power law exponent corresponds to the number of photons needed to overcome the effective barrier height. With increasing intensity ($\gamma \ll 1$), the exponent decreases and the power law approaches linearity as the contribution from tunneling emission becomes dominant. In a double-logarithmic representation, the yield thus appears as a straight line that smoothly bends down to an increasingly smaller slope with increasing intensity.

\begin{figure*}
  \includegraphics[width=0.9\textwidth]{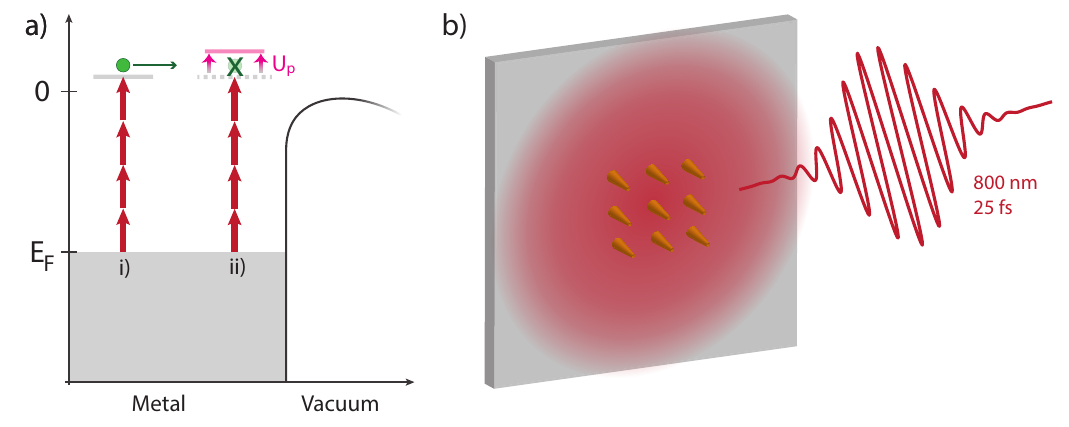}
  \caption{Photoemission models and experimental setup. a) Schematics of two photoemission mechanisms: i) Multiphoton photoemission: An electron at the Fermi level $E_F$ absorbs four photons, enabling it to overcome the effective barrier height and be emitted into the continuum. ii) Channel closing: The intense laser field causes an upshift of the vacuum level by the ponderomotive energy $U_p$ of the electron in the laser field. If the field strength, and thus $U_p$, is high enough, four photons are no longer sufficient to emit an electron, resulting in a decrease (or stagnation) in the emission rate. b) Sketch of the experimental setup (not to scale). A 3x3 square array of sharp gold needle tips with 200 nm pitch is illuminated by ultrashort laser pulses at a central wavelength of \SI{800}{\nano\meter} and a pulse duration of \SI{25}{\femto\second}.}
  \label{fig:MPemission_intro}
\end{figure*}
\noindent These emission models further predict that the electron yield shows a periodic modulation with intensity due to an effect known as \textit{channel closing} \cite{Kopold2002,Yalunin2011}. In a simplified, perturbative picture, the time-dependent ponderomotive potential of the laser pulse leads to an upshift of the vacuum level, such that an $n$-photon process is no longer sufficient to liberate an electron into the vacuum (see Fig.\ref{fig:MPemission_intro}a)). This results in a stagnation or even decrease of the emission rate with increasing intensity, until the yield of the $(n+1)$-photon emission channel becomes large enough to increase the rate again. Consequently, the intensity difference between two neighboring channel closings corresponds to the increase in ponderomotive energy $U_\mathrm{P}$ equal to one photon energy $E_\mathrm{Ph}$.\\
The presence of these ponderomotive shifts was shown both for gas-phase \cite{Muller1988} and needle tip experiments \cite{Schenk2010} by observing energy shifts of above-threshold-emission peaks in spectrally resolved measurements.
Surprisingly, despite the long history of strong-field physics, the expected modulation of the emission rate has so far only been observed in gas-phase systems \cite{Bucksbaum1990,Lai2017,Zimmermann2017}, but not in solid-state systems.\\
Observing these modulations would be of great interest for practical applications of ultrafast photoemission, such as ultrafast electron sources \cite{Hobbs2014,Swanwick2014,Pettine2020,Brueckner2024,Schroeder2025,Chen2025} or petahertz electronics \cite{Ludwig2020,Bionta2021,Arai2023,Heide2024,Davidovich2025}. For the characterization of such devices, determining the emission mechanism and the local field strength at the emission site is of highest relevance. Because optical field enhancement is determined by nanometric and even atomic-scale surface features, the optical nearfield is notoriously difficult to predict and to measure \cite{Thomas2013,Thomas2015}. In the case of single emitters such as needle tips, it is possible to extract information about the local field strength and the field enhancement factor through the measurement of high-energy rescattering cut-offs in photoelectron energy spectra \cite{Schenk2010,Thomas2013,Racz2017,Paschen2023a}. In principle, information about the leading emission process can also be inferred from the intensity scaling of the total photoelectron yield \cite{Heimerl2025JVST}. The theoretically expected bending of the rate towards linearity with increasing intensity has been previously brought forward to explain observed decreases in the scaling of the emission rate from needle tips \cite{Bormann2010,Swanwick2014,Hobbs2014,Keathley2017,Heimerl2025JVST}.\\
For more complex devices however, which can contain thousands of emitters or optoelectronic circuits, such investigations of electron emission processes are considerably more challenging. The total emission current of these devices can reach millions of electrons per laser pulse \cite{Swanwick2014}, saturating many commonly used detectors like microchannel plates or delay-line detectors \cite{Jagutzki2002}. Moreover, the emitted charge per pulse can exceed hundreds of electrons for each emitter, inducing severe space charge effects affecting the emission \cite{Jensen2012,Schoetz2021,Paschen2023}. Crucially, space charge-dominated emission also scales linearly \cite{Paschen2023} with intensity, making any interpretation of the emission behavior solely based on the emission rate ambiguous.\\
Although numerical simulations can estimate field enhancement for known geometries, precisely characterizing the emitter geometry \textit{in situ} remains difficult, and minor sub-nanometer variations on the surface can drastically alter the field enhancement \cite{Thomas2015,Paschen2023a}. However, observing channel closings would make it possible to directly determine the local field strength by matching the observed features to emission models or simulations.\\
Here, we study the strong-field electron emission rate from an array of gold needle tips. We demonstrate the appearance of channel closings in photoemission from a needle tip array and show that these features can serve as a precision gauge for the optical near-field and the field enhancement factor at the emitter apices, without the need for electron energy spectroscopy.\\

\noindent In the experiment, we investigate the non-linear photoemission from a 3x3 square array of gold needle tips with a pitch of 200 nm (Fig.~\ref{fig:MPemission_intro}b)) \cite{Brueckner2024}. The tips have a radius of curvature at the apex of around \SI{8}{\nano\meter}, characterized by scanning electron microscope imaging.
The measurements were carried out in an ultra-high vacuum chamber at a base pressure of $1\cdot10^{-10}$\,hPa. The array was biased at $-200$\,V and the total emission current was recorded using a source measurement unit connected to the tips. Electron emission was triggered using ultrashort laser pulses with a pulse duration of \SI{25}{\femto\second} from a titanium:sapphire oscillator operating at a central wavelength of \SI{800}{\nano\meter} with a repetition rate of 80 MHz and a maximum pulse energy at the sample of $0.84$\,nJ. The pulses were incident at an angle of $7^{\circ}$ and were focused to a spot of 2\,$\mu$m ($1/e^2$ radius) with an off-axis parabolic mirror.

\begin{figure}
  \includegraphics[scale=0.5]{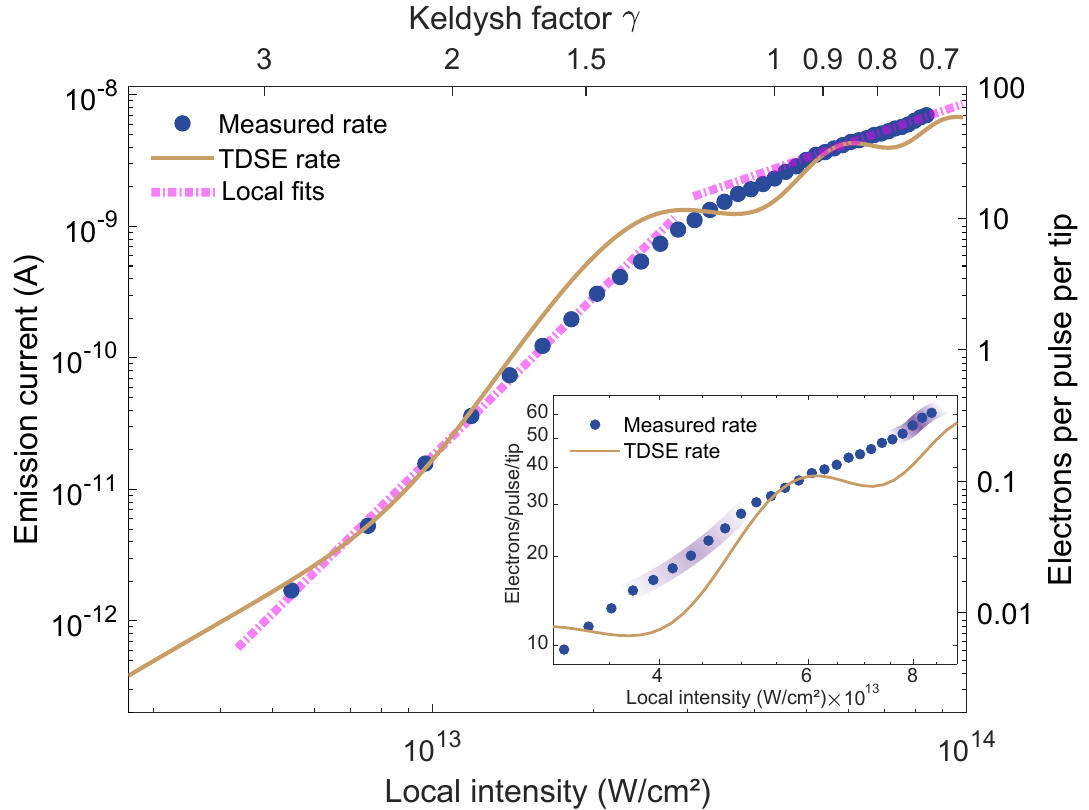}
  \caption{Ultrafast laser-emitted current from the 3x3 tip array (blue dots) plotted against the local peak intensity (bottom axis) and the Keldysh parameter $\gamma$ (top axis). The overall trend indicates a transition from multiphoton to tunnelling emission regime. The pink dashed lines are local fits with slopes of $n=4.0$ and $n=1.4$ representing the intensity scaling at low and high intensities, respectively. Intriguingly, slight oscillations can just be seen on top of the gradual slope decrease. TDSE simulation results (orange line) show these oscillations much clearer. We will show that these oscillations result from channel closing.  See text for details. The intensity axis is calibrated using the procedure detailed below. Inset: Detailed view of the high intensity region, highlighting two channel closing features in the data (shaded in purple as a guide for the eye). The origin of the reduced experimentally observed modulation depth relative to the simulation is discussed in the text.}
  \label{fig:curre_vs_inten}
\end{figure}

\noindent Fig.~\ref{fig:curre_vs_inten} shows the total measured photocurrent (blue dots) versus the local peak intensity of the optical near field at the tips. The local intensity is calculated by scaling the incident intensity with a field enhancement factor $\xi$ of $\xi=13.2$, which was determined using the method presented in this paper (see below). Similar to a previous characterization of these sources, the array exhibits a fairly high emission current of up to 7\,nA \cite{Brueckner2024}. Analyzing the scaling of the emission rate with intensity, a nonlinearity of $n = 4.0$ is found at low local intensities around $8\cdot10^{12}$ \unit{W/cm^2}. The slope of the curve decreases gently with increasing intensity down to a non-linearity of $n=1.4$ at high intensities approaching $8\cdot10^{13}$ \unit{W/cm^2}. From this scaling we infer that at low intensities, emission is dominated by a multiphoton process (see also Fig.~\ref{fig:MPemission_intro}), where approximately four photons of energy $E_\mathrm{Ph} =1.55$\,eV are required to lift electrons over the potential barrier (gold work function $\sim$5.2 eV \cite{Kawano2008}). With increasing intensity, the slope decreases as the emission is increasingly dominated by tunneling \cite{Kim2023,Heimerl2025JVST}.\\
By close inspection of Fig.~\ref{fig:curre_vs_inten}, we see that there are ever so slightly visible oscillations superimposed on the slow gradual slope decrease: Looking at the emission rate in more detail (Fig.~\ref{fig:curre_vs_inten}, inset), a slight decrease and subsequent recovery in the rate scaling can be observed starting at a local intensity of $4\cdot10^{13}$ W/cm². A second oscillation in the rate can be seen at an intensity of $7.5\cdot10^{13}$ W/cm².
To determine the origin of these features, we compare the experimental data to a time-dependent Schrödinger equation (TDSE) simulation of the emission rate. We use a single active electron simulation of a bound electron in a box potential, with the potential width chosen to yield a work function of $5.2$\,eV.\
For details of the simulation see \cite{Dienstbier2023}. We calculate the integrated emission yield for peak intensities ranging from $2.8\cdot10^{12}$ to $9.8\cdot10^{13}$ W/cm² (Fig.~\ref{fig:curre_vs_inten}, orange line). The simulation results clearly show pronounced oscillations in the emission rate, which arise from channel closings as we will show below. Compared to the measured data, the rate modulation in the simulation is much stronger, with the rate completely stagnating during a channel closing, whereas in the experimental data the modulation appears only as a slight dip in an overall increasing rate.

\begin{figure}
  \includegraphics[scale=0.5]{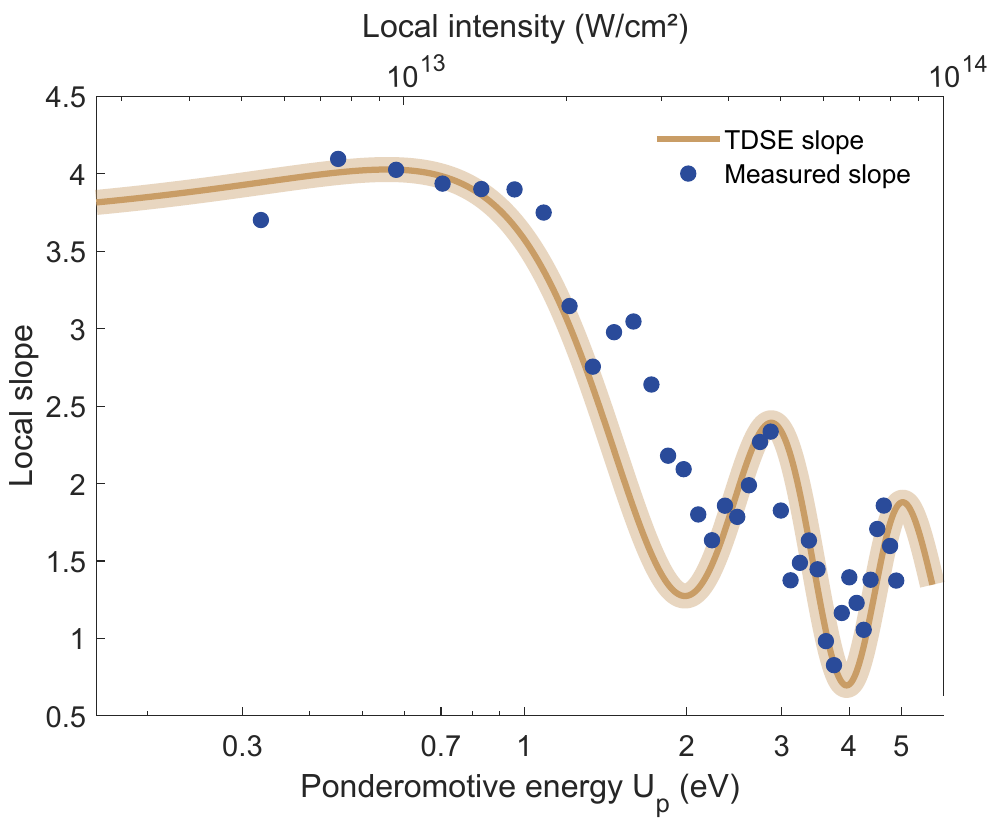}
  \caption{Field enhancement factor calibration via local slope analysis. Local slope of the emitted current (blue dots) plotted against intensity (top axis) and the ponderomotive energy $U_p$ (bottom axis). In this local slope representation, the channel closings are clearly apparent as distinct peaks superimposed on the smoothly decreasing background nonlinearity. Matching these peaks to the peaks present in the slope of the simulated rate (orange line) enables us to precisely determine the local enhanced nearfield intensity and thus the optical field enhancement factor to be $\xi=13.2\pm 0.4$. To indicate the precision of this method, the shaded band around the TDSE curve shows the simulation results with a field enhancement factor of $13.2\pm 0.4$.}
  \label{fig:Slope_pondEnergy}
\end{figure}
\noindent To increase the contrast of the oscillations in the experimental data, we evaluate the local slope of the emission rate using the central difference method with a step size of 2 data points. The results are shown in Fig.~\ref{fig:Slope_pondEnergy} (blue dots). In this local slope representation, the rate oscillations can be clearly distinguished as peaks at $4.9\cdot10^{13}$ W/cm² and $8.4\cdot10^{13}$ W/cm², superimposed on a smoothly decreasing nonlinearity. The local slope extracted from the TDSE simulation shows corresponding peaks. By smoothing over the TDSE data, i.e., by artificially decreasing the contrast, the measured and simulated slopes show remarkable agreement. We confirmed that this smoothing only affects the amplitude of the peaks, but not the spacing in intensity. Physically, the smoothing corresponds to local intensity variations across the tip array caused by minor variations of individual tip geometries.
The clear matching of the peaks of data and simulation allows us to calibrate the local intensity at the tip array \textit{in situ}. To achieve quantitative agreement, the incident peak field in the experiment requires scaling by a field enhancement factor of $\xi=13.2$. The resulting local peak intensity at the tip apices hence reaches up to $8.5\cdot10^{13}$ \unit{W/cm^2}. Consequently, for most of the measured intensity range $\gamma\lesssim1$, confirming that the emission is dominated by non-adiabatic tunneling \cite{Ivanov2005}. We can, in addition, infer that the emission does not seem to be significantly influenced by space charge effects, which, for the intensities and emission currents reached in this experiment, is in line with previous investigations \cite{Schoetz2021}.

\begin{figure}
  \includegraphics[scale=0.53]{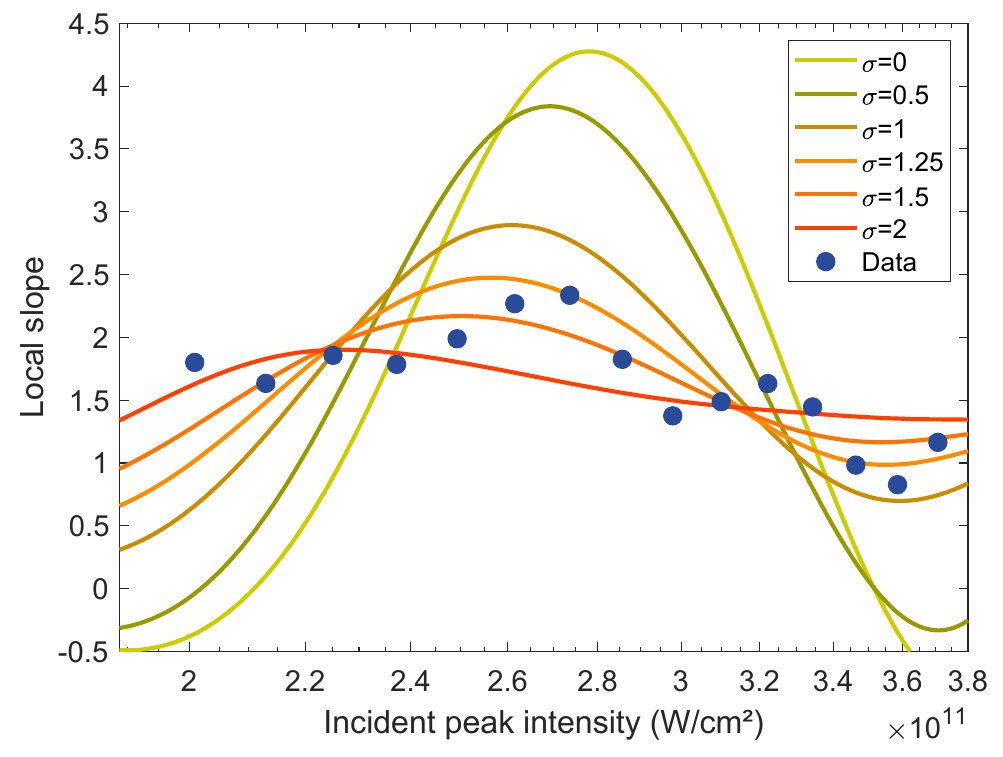}
  \caption{Loss of contrast due to geometric tip variations. Local slope of the experimental data (blue dots) and the simulated emission rates against the incident peak intensity around the first observed channel closing. The field enhancement factor is modeled as a Gaussian distribution with width $\sigma$ (yellow: $\sigma = 0$, increasing in steps of $0.5$). The red curve with $\sigma = 2$ is clearly too shallow. Comparable peak height is achieved at $\sigma_\textrm{opt} \approx 1.25$, meaning that the field enhancement factor varies across the array from 11.95 to 14.45.}
  \label{fig:LocalSlope_IncidentPeakInt}
\end{figure}
\noindent The contrast, i.e. the modulation depth in the experimental data, is significantly smaller than in the single-tip TDSE simulations. As stated before, we were able to reproduce this contrast reduction by strongly smoothing over the simulation data. The origin of this loss of contrast might lie in variations of the geometry, and thus the field enhancement factor, between the different tips in the array. In the following, we will attempt to reproduce this effect using a simple model. We model the geometric variation of the tips as a Gaussian distribution $\frac{1}{\sigma \sqrt{2\pi}}\mathrm{exp}\left( -\frac{(\xi_{\mathrm{var}}-\xi)^2}{2\sigma^2}\right)$ of the field enhancement factor $\xi_{\mathrm{var}}$ with a width $\sigma$ around the determined mean value of $\xi = 13.2$. We calculate the emission rate based on the simulated yield for different field enhancement factors from 11 to 16. These yields are weighted with the Gaussian distribution, summed up and the corresponding local slope of the total yield is calculated. The results are plotted against the incident, i.e. non-enhanced, laser peak intensity in Fig.~\ref{fig:LocalSlope_IncidentPeakInt}, together with the first peak in the measured data, which appears at a local peak intensity of about $4.8\cdot10^{13}$ W/cm² and a ponderomotive energy of 2.9 eV. We observe that the peak becomes increasingly lower and less pronounced for increasing $\sigma$. This confirms that variations in the tip geometries can be a significant contribution to the washing out of the modulation. By comparing the simulated peak heights to the experimental data, we estimate the variation in the field enhancement factor to be around $\sigma=1.25$. Field enhancement simulations \cite{Thomas2013} indicate that this spread corresponds to tip radii variations of roughly 1 nm around a mean of 11 nm, assuming all other geometry parameters remain constant. 
This could indicate that a slight blunting and reforming of the tip apices has taken place under the influence of the high laser peak fields.
\\
\noindent The high agreement between our data and the TDSE simulation allows us to  make two further interesting observations. 
First, we find that the energy spacing between subsequent channel closings is roughly 2\,eV, which exceeds the photon energy of $1.55$\,eV expected from a simple multiphoton emission model. Most likely, this discrepancy arises because needle tips, unlike atoms, have a strong optical near field, which decays over a short distance on the order of the tip size, i.e. a few nanometers. This decay can affect the quivering of the electron in the field and, thus, the ponderomotive energy experienced by an electron \cite{Herink2012,Heimerl2025}. Second, we observe that channel closings appear at Keldysh parameters $\gamma < 1$, where the emission is dominated by tunneling. Here, the simple interpretation of a perturbative multiphoton process affected by a shift in the vacuum level is no longer valid. A different model would be needed for a cohesive explanation of this process, which is beyond the scope of this work.

\noindent Finally, we note that at the highest peak intensities, the emission exceeds 50 electrons per pulse per tip in the array, indicating that space charge effects could potentially affect the emission behavior. Over the duration of the laser pulse, the cloud of emitted electrons around the tip apex can partially shield the tip from the incoming laser field, leading to a linear scaling \cite{Jensen2012,Schoetz2021,Paschen2023}. Hence, the mere observation of a close-to-linear scaling in photoemission does not allow us to conclude that the emission is in the tunneling regime and that the observed features originate from strong-field effects. Rather, it is the additional channel closing features that only allow us to extract the local intensity precisely.\\

\noindent To summarize, we found clear signatures of channel closing in ultrafast photoemission currents from an array of sharp gold needle tips, not observed before from a solid state emitter. By matching channel closing peaks in the photocurrent slope to a TDSE simulation of the rate, we directly determined the local field enhancement factor at the emission sites to $\xi = 13.2\pm 0.4$, as well as the emitter tip radii uniformity. In essence, from the simple measurement of photocurrent versus laser intensity we were able to draw clear, unambiguous conclusions about the emission process and the local optical field at the nanometric sharp tips including field enhancement. This method does not require any additional devices such as electron spectrometers or other complex and expensive detectors, making it easily applicable to any ultrafast photoemission experiment. In particular, this method could be useful especially for photoemission-based petahertz electronics, where well-established near-field characterization techniques may be impractical or inapplicable.\\
\\

\textit{Acknowledgements}---The authors thank Stefan Meier for helpful discussions. This research was supported by the Gordon and Betty Moore Foundation (Grant: Imaging Quantum Coherence with Shaped Electrons (iQCE)), the
European Research Council (Advanced Grant AccelOnChip), the BMBF (Grant DLA e-prep) and the Deutsche Forschungsgemeinschaft (DFG, German
Research Foundation) through TRR 306 QuCoLiMa (‘Quantum
Cooperativity of Light and Matter’).

\textit{Data availability}---The data used for this work are available from the authors upon reasonable request.

\bibliography{bibliography}% Produces the bibliography via BibTeX.

\end{document}